\documentclass{iopart}
\usepackage{graphicx}
\usepackage{bm}

\newcommand{\hK}{\hat{K}}
\newcommand{\hU}{\hat{U}}
\newcommand{\hV}{\hat{V}}
\newcommand{\hT}{\hat{\Theta}}
\newcommand{\hP}{\hat{P}}
\newcommand{\hb}{\hat{\beta}}
\newcommand{\hd}{\hat{\delta}}
\newcommand{\hTT}{\hat{T}}

\newcommand{\hHD}{\hat{\cal H}}
\newcommand{\hHDp}{\hat{\cal H}^{'}}

\newcommand{\bRho}{\bar{\rho}}

\begin{document}

\title[Non-locality of transformations in a magnetic field]
      {Non-locality of energy separating transformations for Dirac electrons in a magnetic field}
\date{\today}
\author{Tomasz M Rusin$^1$ and Wlodek Zawadzki$^2$}
\address{$^1$ Orange Customer Service sp. z o. o., ul. Twarda 18, 00-105 Warsaw, Poland\\
         $^2$Institute of Physics, Polish Academy of Sciences, 02-668 Warsaw, Poland}
 \ead{Tomasz.Rusin@orange.com}

\submitto{\JPA}

\begin{abstract}
We investigate a non-locality of Moss-Okninski transformation~(MOT) used to separate positive
and negative energy states in the 3+1 Dirac equation for relativistic
electrons in the presence of a magnetic field.
Properties of functional kernels generated by the MOT are analyzed and kernel
non-localities are characterized by calculating their second moments parallel and perpendicular to
the magnetic field. Transformed functions are described and investigated by computing their variances.
It is shown that the non-locality of the energy-separating transformation in the direction parallel
to the magnetic field is characterized by the Compton wavelength~$\lambda_c=\hbar/mc$. In the
plane transverse to magnetic field the non-locality depends both on magnetic radius~$L=(\hbar/eB)^{1/2}$
and~$\lambda_c$. The non-locality of MO transformation for the 2+1 Dirac equation is also considered.
\end{abstract}

\maketitle
\section{Introduction}
The Dirac equation, in spite of its numerous applications and fundamental significance in
relativity and quantum mechanics, is far from being completely understood.
This is especially true for the Dirac electrons in the presence of external fields. For example,
it has been known for many years that the Dirac equation can be solved exactly in case of an
external uniform magnetic field~\cite{Rabi1928}, but it is only recently that an
explicit electron dynamics of Dirac electrons in the presence of a magnetic field
was worked out~\cite{Rusin2010}. This problem is related to a somewhat mysterious
phenomenon known in the literature under the name of Zitterbewegung.
The phenomenon was conceived in 1930 by Schrodinger who observed that, according to
the Dirac equation, operators corresponding to the electron velocity do not commute
with the Dirac Hamiltonian, so the velocity depends on time also in absence of
external fields~\cite{Schrodinger1930}. It is understood at present that this remarkable
phenomenon is due to an
interference of states corresponding to positive and negative electron
energies~\cite{BjorkenBook,SakuraiBook,GreinerBook}. In fact, it is known by
now that, whenever one deals with a spectrum of positive and negative energy branches,
the Zitterbewegung will appear~\cite{Zawadzki2011}.

There exist ways to avoid the duality of positive and negative energies. The best known
of them is the Foldy-Wouthuysen transformation~(FWT) which, in case of no external
fields, allows one to break exactly the~$4\times 4$ Dirac equation into
two~$2\times 2$ equations for positive or negative energies~\cite{Foldy1950}. In their
pioneering paper, FW remarked that a kernel transforming functions from
the original to the FW representation is characterized by a
non-locality in coordinate space on the order of the Compton
wavelength~$\lambda_c=\hbar/mc$. Rose in his book~\cite{RoseBook} put this statement on a quantitative
basis by showing that the second moment of the functional kernel is equal to~$(3/4)\lambda_c^2$.
This result was not followed by other
investigations and it is by now not well known. The problem was recently taken up
by the present authors who demonstrated that also other transformed quantities
are characterized by non-localities on the order of~$\lambda_c$~\cite{Rusin2011}.

Case~\cite{Case1954} showed that also the Dirac equation including the presence
of an external uniform magnetic field~${\bm B}$ can be exactly separated into
two equations for positive and negative electron energies. This form allows one
to find easily the energy eigenvalues in any desired semi-relativistic approximation
(see our Eq.~(\ref{DK_UHU}) and Ref.~\cite{Zawadzki2005AJP}).
It is possible to separate the positive and negative
electron energies for~$B\neq 0$ because the presence of a uniform magnetic field does not induce
electron transitions between the two parts of the spectrum.
The case of magnetic field is of interest for several reasons. First, it
introduces an adjustable external parameter~$B$ affecting all energies and other
electron characteristics. Second, it quantizes the orbital motion in
two degrees of freedom. Third, it quantizes the spin energies. However, to our knowledge properties
of the separating transformations for this important case were not investigated until now.

In the form proposed by Case the transformation operator is~$\hU=e^{i\hat{\rho}_2\phi/2}$
with~$\hat{\rho}_2$ being the standard Dirac matrix, and the phase
given by~$\phi=\tan^{-1}(\hat{\bm \sigma}\hat{\bm \pi})/mc$, where~$\hat{\bm \sigma}$
are the Pauli matrices and~$\hat{\bm \pi}$ is the generalized momentum operator including the
vector potential of the magnetic field. In this form it is difficult to analyze the
properties of~$\hU$. A method to get a more explicit form of~$\hU$ was described by
de Vries~\cite{deVries1970}. However,
it was shown that the ''separating`` transformations, both in absence
of fields and for~${\bm B} \neq 0$, are not unique. In other words, it is possible to separate
the Dirac equation into two equations for positive or negative energies, both in
absence of fields and in the presence of a magnetic field, using {\it different} unitary
transformations~\cite{Case1954,deVries1970,Tsai1973,Weaver1975,Moss1976}.
In our description involving Dirac electrons in a magnetic field we use a two-step
transformation proposed by Moss and Okninski~(MO)~\cite{Moss1976}. An advantage of this
choice is that the MOT is given explicitly, it is convenient for kernel
calculations and we can compare our final results
in the limit~$B\rightarrow 0$ with our previous description for~$B=0$, see~\cite{Rusin2011}.
In our treatment we do not concentrate so much on the separation of energies but rather on
the properties of transformation operators and the transformed functions.
Naturally, we emphasize the effects of an external magnetic field.

Our subject is of relevance for at least three reasons. First, with the comeback of
interest in narrow-gap semiconductors and, in particular, the appearance of zero-gap
graphene~\cite{Novoselov2004}, there has been a real surge of works concerned with
relativistic-type wave equations. Second, it is now possible
to simulate the Dirac equation with the trapped ions or cold atoms interacting with
laser radiation, where one can tailor ''user friendly`` values of the basic
parameters~$mc^2$ and~$c$~\cite{Rusin2010,Leibfried2003,Lamata2007,Johanning2009}.
In fact, a proof-of-principle experiment simulating the 1+1 Dirac equation and the
resulting Zitterbewegung was carried out by Gerritsma {\it et al.}~\cite{Gerritsma2010}.
Third, one can simulate equally well the Dirac equation in the presence of
a magnetic field~\cite{Rusin2010,Lamata2011}. In such simulations it is possible to tailor a
seeming intensity of the field making its effects on relativistic electrons
relevant in terrestrial conditions. Putting it in more quantitative terms, one
can achieve experimentally the so-called Schwinger field,
defined by the equality~$\hbar eB_0/m=mc^2$, not only in the vicinity of neutron stars
but in standard laboratories, see our discussion below.

Our paper is organized in the following way. In Sections II and III we introduce the
Moss-Okninski transformation, describe its functional kernel and calculate its zeroth and
second moments. In Section IV we transform Gaussian functions and calculate their normalized
variances. In Section V we analyze kernels in two dimensions.
In Section VI we discuss our results. The paper is concluded by a summary. In two
Appendices we give some mathematical details of our derivations.

\section{Transformation Kernel and its non-locality}

In this section we define the MO transformation and its functional kernel. Then
we analyze kernel's properties and, in particular, its non-locality.
The Dirac Hamiltonian for a relativistic electron in a magnetic field is
\begin{equation} \label{DK_HD}
\hHD = c\hat{\alpha}_x\hat{\pi}_x + c\hat{\alpha}_y\hat{\pi}_y + c\hat{\alpha}_z\hat{\pi}_z + \hb mc^2,
\end{equation}
where~${\hat{\bm \pi}} = {\hat{\bm p}}+e{\bm A}$ is the generalized momentum,~$|e|$ is the electron
charge,~${\bm A}$ is the vector potential of a magnetic field~${\bm B}$, and~$\hat{\alpha}_i$
and~$\hb$ are the Dirac matrices in the standard notation. Following Moss and Okninski~\cite{Moss1976}
we first introduce a unitary operator
\begin{equation} \label{DK_V} \hV=\hd(\hd+\hb)/\sqrt{2},\end{equation}
in which
\begin{equation} \label{DK_hd}
 \hd=\hat{\alpha}_x\hat{\alpha}_y\hat{\alpha}_z\hb =
 \left(\begin{array}{cccc} 0 & 0 &-i & 0\\ 0 & 0 & 0 & -i\\
  i & 0 & 0 & 0 \\ 0 & i & 0 &0 \end{array}\right).\end{equation}
This operator transforms the initial Hamiltonian into a form with vanishing diagonal terms
\begin{equation} \label{DK_HDp}
\hHDp=\hV\hHD\hV^{\dagger} =
 c\hat{\alpha}_x\hat{\pi}_x + c\hat{\alpha}_y\hat{\pi}_y + c\hat{\alpha}_z\hat{\pi}_z + \hd mc^2.
\end{equation}
The second step transforms~$\hHDp$ to a strictly diagonal form with the use of an operator~$\hU$
\begin{equation} \label{DK_U}
 \hU = \frac{1}{\sqrt{2}} \left(\hb + \hT \right),
\end{equation}
where
\begin{equation} \label{DK_T}
 \hT = \frac{\hHDp}{(\hHD^{'2})^{1/2}}.
\end{equation}
After the complete transformation the Hamiltonian has a diagonal form
(cf. Refs.\cite{Case1954,deVries1970,Moss1976,Zawadzki2005AJP})
\begin{equation}\label{DK_UHU}
\hU\hHDp \hU^{\dagger}=\beta(m^2c^4 + c^2\hat{\bm \pi}^2 -ec^2\hbar \hat{\bm \Sigma}{\bm B})^{1/2},
\end{equation}
where~$\hat{\bm \Sigma}=\left(\begin{array}{cc} {\bm \sigma} & 0 \\ 0 & {\bm \sigma} \end{array}\right)$
is the~$4\times 4$ spin operator.
Since~$\hat{\bm \pi}$ is an operator, the square root is understood as an
expansion in powers of~$\hat{\bm \pi}$. In the following we concentrate on the
properties of the transformation and transformed functions.

Let~$|\Psi\rangle$ be a function in the old space and~$|\Psi'\rangle=\hV|\Psi\rangle$ in the transformed space.
We introduce standard projecting operators for positive and negative energy states
\begin{equation} \label{DK_P}
\hP^{\pm}=\frac{1}{2}(\hat{1} \pm \hT).
\end{equation}
If the projection operators are applied to the transformed functions~$|\Psi'\rangle$,
they split them into states
corresponding to positive or negative energies:~$|\Psi'^{\pm}\rangle = \hP^{\pm}|\Psi'\rangle$.
Next~$|\Psi'^{\pm}\rangle$ are transformed using the operator~$\hU$,
i.e.~$|\tilde{\Psi}'^{\pm}\rangle=\hU|\Psi'^{\pm}\rangle$. We introduce
operators~$\hU^{\pm}$ defined as~$\hU^{\pm}=\hU\hP^{\pm}$ with the
property~$|\tilde{\Psi}'^{\pm}\rangle=\hU\hP^{\pm}|\Psi'\rangle$. We have
\begin{eqnarray} \label{DK_Upm}
 \hU^{\pm} &=& \frac{1}{2\sqrt{2}} \left(\hb + \hT \right) \left(\hat{1} \pm \hT\right) \nonumber \\
           &=& \frac{1}{2\sqrt{2}} \left(\hb \pm \hat{1}\right) \left(\hat{1} +\hT \right).
\end{eqnarray}
Let us consider transformed functions~$\tilde{\Psi}'^{\pm}({\bm r}_1) = \langle{\bm r}_1|\hU^{\pm}|\Psi'\rangle$. Inserting the identity operator~$\hat{1}=\int|{\bm r}_2\rangle\langle {\bm r}_2| d^3{\bm r}_2$ one has
\begin{eqnarray} \label{DK_Psi_pm_r}
 \tilde{\Psi}'^{\pm}({\bm r}_1)
  &=& \int \langle {\bm r}_1 |\hU^{\pm}| {\bm r}_2\rangle
       \langle {\bm r}_2 |\Psi'\rangle d^3{\bm r}_2 \nonumber \\
  &=& \int \hK^{\pm}( {\bm r}_1, {\bm r}_2) \Psi'( {\bm r}_2) d^3{\bm r}_2,
\end{eqnarray}
where we defined the functional transformation kernels
\begin{equation} \label{DK_Kpm}
\hK^{\pm}( {\bm r}_1, {\bm r}_2) = \langle {\bm r}_1 |\hU^{\pm}| {\bm r}_2\rangle.
\end{equation}

In the following we investigate the properties of~$\hK^{\pm}( {\bm r}_1, {\bm r}_2)$.
To calculate~$\hK^{\pm}({\bm r}_1, {\bm r}_2)$ we insert twice
the unity operator~$\hat{1}=\sum_{\rm m}|{\rm m}\rangle\langle {\rm m}|$.
The states~$|{\rm m}\rangle$ are eigenstates of the Hamiltonian~$\hHDp$ given in Eq.~(\ref{DK_HDp}),
and the summation is carried out over all quantum
numbers characterizing~$|{\rm m}\rangle$. Thus
\begin{equation} \label{DK_K12}
\hK^{\pm}({\bm r}_1,{\bm r}_2) =
   \sum_{\rm m,m'}\langle {\bm r}_1|{\rm m}\rangle \langle {\rm m}|\hU^{\pm}|{\rm m'} \rangle
   \langle {\rm m'}|{\bm r}_2\rangle.
\end{equation}
The operator~$\hU^{\pm}$ consists of operators:~$\hat{1}$,~$\hb$ and~$\hT$, see Eq.~(\ref{DK_Upm}).
The matrix elements of~$\hb$ and~$\hat{1}$
are diagonal:~$\hb_{\rm m,m'}=\hb\delta_{\rm m,m'}$ and~$\hat{1}\delta_{\rm m,m'}$, respectively.
The non-diagonal contribution to~$\hK^{\pm}({\bm r}_1,{\bm r}_2)$ arises from~$\hT$
\begin{equation} \label{DK_T12}
\hT({\bm r}_1,{\bm r}_2) =
   \sum_{\rm m,m'}\langle {\bm r}_1|{\rm m}\rangle \langle {\rm m}|\hT|{\rm m'} \rangle
    \langle {\rm m'}|{\bm r}_2\rangle.
\end{equation}
The eigenstates~$|{\rm m}\rangle$ of the Hamiltonian~$\hHDp$ are related to the
eigenstates~$|{\rm n}\rangle$ of the initial Dirac Hamiltonian~$\hHD$ by a unitary
transformation
\begin{equation} \label{DK_mVn}|{\rm m}\rangle=\hV|{\rm n}\rangle, \end{equation}
where~$\hV$ is defined in Eq.~(\ref{DK_V}).
We want to calculate the elements of~$\hT({\bm r}_1,{\bm r}_2)$ matrix and for this purpose
we need to know explicit forms of the eigenstates~$|{\rm m}\rangle$.

We take the vector potential of a magnetic field in the
Landau gauge~${\bm A}=(-By,0,0)$. Then the eigenstates~$|{\rm n}\rangle$ depend on
five quantum numbers:~$n,k_x,k_z,\epsilon,s$, where~$n$ is the harmonic oscillator
number,~$k_x$ and~$k_z$ are wave vector components,~$\epsilon=\pm 1$ labels positive
and negative energy branches, and~$s=\pm 1$ is the spin index
corresponding to the spins~$\pm 1/2$, respectively.
In the Johnson-Lippman representation~\cite{Johnson1949} the
eigenfunctions~$\Phi_{\rm n}({\bm r})=\langle {\bm r}|{\rm n}\rangle$ of the Hamiltonian~$\hHD$ are
\begin{equation} \label{Ei_Lippman}
\Phi_{\rm n}({\bm r}) = N_{n\epsilon p_z}\left(\begin{array}{rl}
   s_1(\epsilon E_{n,k_z}+mc^2)  &\phi_{n-1}({\bm r})\\
   s_2(\epsilon E_{n,k_z}+mc^2)  &\phi_{n}  ({\bm r})\\
  (s_1cp_z -s_2\hbar \omega_n)   &\phi_{n-1}({\bm r})\\
  -(s_1\hbar \omega_n + s_2 cp_z)&\phi_{n}  ({\bm r}) \end{array}\right),
\end{equation}
where~$s_1=(1+s)/2$ and~$s_2=(1-s)/2$ select the states~$s=\pm 1$, respectively.
The frequency is~$\omega_n=\omega \sqrt{n}$ with~$\omega=\sqrt{2}c/L$, the energy is
\begin{equation} \label{Ei_Enkz}
 E_{n,k_z}=\sqrt{(mc^2)^2 + (\hbar\omega_n)^2 + (\hbar k_zc)^2 },
\end{equation}
and it does not depend explicitly on~$s$.
The norm in Eq.~(\ref{Ei_Lippman}) is~$N_{n\epsilon k_z}=(2E_{n,k_z}^2+2\epsilon mc^2E_{n,k_z})^{-1/2}$,
and the~$\phi_{n}$ functions are
\begin{equation} \label{Ei_phir}
 \phi_{n}({\bm r}) = \frac{e^{ik_xx+ik_zz}}{2\pi\sqrt{L}C_n}{\rm H}_{n}(\xi)e^{-1/2\xi^2},
\end{equation}
where~${\rm H}_{n}(\xi)$ are the Hermite polynomials,~$C_n=\sqrt{2^n n!\sqrt{\pi}}$,
the magnetic radius is~$L=\sqrt{\hbar/eB}$ and~$\xi=y/L-k_xL$.

Since~$\hV$ defined above is an unitary matrix, there is~$E_{\rm m}=E_{\rm n}=\epsilon E_{n,k_z}$.
The states~$|{\rm m}\rangle$ are characterized by the same five quantum
numbers describing~$|{\rm n}\rangle$, so
there is also~$\delta_{\rm m,m'} =\delta_{\rm n,n'}$.
For two eigenstates~$|{\rm m}\rangle$ and~$|{\rm m'}\rangle$ of the Hamiltonian~$\hHDp$ we have:
$\langle {\rm m}|\hHDp|{\rm m'}\rangle=E_{\rm m}\delta_{\rm m,m'}$,
$\langle {\rm m}|\hHD^{'2}|{\rm m'}\rangle= E_{\rm m}^2\delta_{\rm m,m'}$, and
\begin{equation} \label{T_Ta}
 \langle {\rm m}|\hT|{\rm m'}\rangle = \frac{E_{\rm m}}{\sqrt{E_{\rm m}^2}}\ \delta_{\rm m,m'}
    = q\epsilon \delta_{\rm n,n'},
\end{equation}
where~$q=1$ or~$q=-1$. Selecting~$q=1$ we obtain
from Eqs.~(\ref{DK_T12}), (\ref{DK_mVn}) and~(\ref{T_Ta})
\begin{equation} \label{T_Tb}
\langle {\bm r}_1|\hT|{\bm r}_2\rangle = \sum_{\rm m} \epsilon
\langle {\bm r}_1|{\rm m} \rangle \langle {\rm m}|{\bm r}_2 \rangle =
\sum_{\rm n} \epsilon \langle {\bm r}_1|\hV{\rm n} \rangle \langle {\rm n}\hV|{\bm r}_2 \rangle.
\end{equation}
Selecting~$q=-1$ in Eq.~(\ref{T_Ta}) would not have changed the results because
we perform a summation over~$\epsilon=\pm 1$. Since~$\hV$ is a number matrix,
there is~$\langle {\bm r}|\hV{\rm n} \rangle=\hV \langle {\bm r}|{\rm n} \rangle$, and one
can rewrite~$\langle {\bm r}_1|\hT|{\bm r}_2\rangle$ in the form
\begin{equation} \label{T_T0}
 \langle {\bm r}_1|\hT|{\bm r}_2\rangle = \hV \hT_0({\bm r}_1,{\bm r}_2) \hV^{\dagger},
\end{equation}
where we introduced an operator~$\hT_0$ by the relation
\begin{equation} \label{T_T0b}
\langle {\bm r}_1|\hT_0|{\bm r}_2\rangle =
\sum_{\rm n} \epsilon \langle {\bm r}_1|{\rm n} \rangle \langle {\rm n}|{\bm r}_2 \rangle.
\end{equation}
The advantage of~$\hT_0({\bm r}_1,{\bm r}_2)$, as compared to~$\hT({\bm r}_1,{\bm r}_2)$,
is that~$\hT_0({\bm r}_1,{\bm r}_2)$ can be directly expressed in terms of the
eigenstates~$|{\rm n}\rangle$ given by Eq.~(\ref{Ei_Lippman}).
Writing in Eq.~(\ref{T_T0b}) all quantum numbers we have
\begin{equation} \label{T_T012}
\langle {\bm r}_1|\hT_0|{\bm r}_2\rangle =
 \int_{\infty}^{\infty}\!\! dk_z\int_{\infty}^{\infty}\!\! dk_x \sum_{n,s,\epsilon}
 \epsilon\Phi_{\rm n}({\bm r}_1) \Phi_{\rm n}^{\dagger}({\bm r}_2),
\end{equation}
where~$\Phi_{\rm n}({\bm r})$ are given in Eq.~(\ref{Ei_Lippman}).
The integrals over~$k_x$ and~$k_z$ as well as the summations
over~$\epsilon$ and~$s$ may be done analytically, the summation over~$n$ is to
be performed numerically. However, in the limit of strong magnetic fields analytical
results may be obtained. The summations and integrations indicated in Eq.~(\ref{T_T012})
are carried out in the following order. First,
we sum over the spin variable~$s$ noting that~$s_1s_2=0$ and~$\sum_s s_1^2 =\sum_s s_2^2 = 1$.
In the summation over~$\epsilon$ we use the property:~$\sum_{\epsilon=\pm 1} \epsilon=0$.
As a result, one obtains
\begin{eqnarray} \label{T0_phir}
\hT_0({\bm r}_1,{\bm r}_2) = \sum_{n=0}^{\infty} \int_{-\infty}^{\infty}\!\!\! d k_x \!\!
 \int_{-\infty}^{\infty}\!\!\! dk_z \! \nonumber \\
\hspace*{-8em}
\left(\begin{array}{cccc}
   \gamma_1\phi_{n'}({\bm r}_1)\phi_{n'}^*({\bm r}_2) & 0 & \gamma_3 \phi_{n'}({\bm r}_1)\phi_{n'}^*({\bm r}_2) &
      -\gamma_2\phi_{n'}({\bm r}_1)\phi_{n}^*({\bm r}_2) \\
   0 & \gamma_1\phi_{n}({\bm r}_1)\phi_{n}^*({\bm r}_2) & -\gamma_2 \phi_{n}({\bm r}_1)\phi_{n'}^*({\bm r}_2) &
      -\gamma_3\phi_{n}({\bm r}_1)\phi_{n}^*({\bm r}_2)  \\
   \gamma_3 \phi_{n'}({\bm r}_1)\phi_{n'}^*({\bm r}_2) & -\gamma_2 \phi_{n'}({\bm r}_1)\phi_{n}^*({\bm r}_2) &
       -\gamma_1 \phi_{n'}({\bm r}_1)\phi_{n'}^*({\bm r}_2) & 0 \\
  -\gamma_2 \phi_{n}({\bm r}_1)\phi_{n'}^*({\bm r}_2) & -\gamma_3 \phi_{n}({\bm r}_1)\phi_{n}^*({\bm r}_2) & 0 &
       -\gamma_1 \phi_{n}({\bm r}_1)\phi_{n}^*({\bm r}_2)
    \end{array}\right), \ \ \ \ \
\end{eqnarray}
where~$\gamma_1=mc^2/(2\pi E)$,~$\gamma_2=\hbar\omega_n/(2\pi E)$,~$\gamma_3=cp_z/(2\pi E)$ and~$n'=n-1$.
The matrix elements of~$\hT_0({\bm r}_1,{\bm r}_2)$ are products of integrals over~$k_z$ and~$k_x$.
There appear three different integrals over~$k_z$:~$D_1$,~$D_2$
and~$D_3$ and four integrals over~$k_x$ denoted as~$\Lambda_{m,n}$. We write explicitly
\begin{eqnarray} \label{T0_sumn}
\hspace*{-8em}
\hT_0({\bm r}_1,{\bm r}_2) = \sum_{n=0}^{\infty}
\left(\begin{array}{cccc}
    D_1 \Lambda_{n-1,n-1} & 0 & D_3 \Lambda_{n-1,n-1} & - \sqrt{n} D_2 \Lambda_{n-1,n}  \\
    0 & D_1 \Lambda_{n,n} & - \sqrt{n}D_2 \Lambda_{n,n-1} & - D_3 \Lambda_{n,n}         \\
    D_3 \Lambda_{n-1,n-1} & - \sqrt{n}D_2 \Lambda_{n-1,n} & - D_1 \Lambda_{n-1,n-1} & 0 \\
  - \sqrt{n} D_2 \Lambda_{n,n-1} & - D_3 \Lambda_{n,n} & 0& - D_1 \Lambda_{n,n}
    \end{array}\right).
\end{eqnarray}
The integrals over~$k_x$ are
\begin{equation} \label{T0_Lambda}
\Lambda_{m,n} = \int_{-\infty}^{\infty} \frac{e^{ik_x(x_1-x_2)}}{C_mC_nL}
               {\rm H}_{m}(\xi_1){\rm H}_{n}(\xi_2)e^{-1/2(\xi_1^2+\xi_2^2)} dk_x,
\end{equation}
where~$\xi_j=y_j/L-k_xL$ with~$j=1,2$.
They are calculated analytically
\begin{eqnarray} \label{T0_Lambda_mn}
\Lambda_{n-j,n-j} &=&\frac{1}{L^2}{\rm L}_{n-j}^0(\bRho^2) e^{-\bRho^2/2+i\chi}, \\
                 \label{T0_Lambda_n1n}
\Lambda_{n-1,n} &=& \frac{r_{1,0}}{L^2\sqrt{n}} {\rm L}_{n-1}^1 (\bRho^2) e^{-\bRho^2/2+i\chi}, \\
                 \label{T0_Lambda_nn1}
\Lambda_{n,n-1} &=& \frac{r_{0,1}}{L^2\sqrt{n}} {\rm L}_{n-1}^1 (\bRho^2) e^{-\bRho^2/2+i\chi},
\end{eqnarray}
where~$j=0,1$,~$\bRho = [(x_1-x_2)^2+(y_1-y_2)^2]/(2L^2)$,~${\rm L}_n^{\mu}(\bRho)$ are
the associated Laguerre polynomials,~$r_{1,0}=[(y_2-y_1)-i(x_1-x_2)]/(L\sqrt{2})$
and~$r_{0,1}=[(y_1-y_2)-i(x_1-x_2)]/(L\sqrt{2})$.
The gauge-dependent phase factor is
\begin{equation} \label{T0_chi} \chi=(x_1-x_2)(y_1+y_2)/2L^2. \end{equation}
It is not translational invariant. The integrals over~$k_z$ are
\begin{eqnarray}
\label{T0_D1}
D_1 &=& \frac{mc^2}{2\pi} \int_{-\infty}^{\infty} \frac{e^{ik_zz_{12}}}{E_{n,k_z}}dk_z
     = \frac{1}{\pi \lambda_c} K_0(|\zeta|a_n), \\
\label{T0_D2}
D_2 &=& \frac{\hbar\omega}{2\pi}\int_{-\infty}^{\infty} \frac{e^{ik_zz_{12}}}{E_{n,k_z}}dk_z
     = \frac{\eta}{\pi\lambda_c} K_0(|\zeta|a_n), \\
     \label{T0_D3}
 D_3 &=& \frac{c\hbar}{2\pi} \int_{-\infty}^{\infty} \frac{ k_z e^{ik_zz_{12}}}{E_{n,k_z}}dk_z
    =\frac{i}{\pi\lambda_c} \left( \frac{1}{\zeta} +{\cal A}(a_n\zeta)\!\!\right),\ \
\end{eqnarray}
in which~$z_{12}=z_1-z_2$ and
\begin{equation} \label{T0_an} a_n=\sqrt{1+\eta^2n}. \end{equation}
Further,~$\lambda_c=\hbar/(mc)$ is the Compton
wavelength,~$\zeta=(z_1-z_2)/\lambda_c$,~$\eta=(2\hbar\omega_c/mc^2)^{1/2}$,
the cyclotron frequency is~$\omega_c=eB/m$, and~$K_0(x)$ is the Bessel function of the second
kind. The quantity~${\cal A}(a_n\zeta)$ is
\begin{eqnarray} \label{T0_Anger}
\hspace*{-3em} {\cal A}(a_n\zeta)= \lim_{\nu \rightarrow 1}\frac{\pi}{\sin(\pi\nu)a_n}
 \left[\sin\left(\frac{\nu\pi}{2} \right)I_{\nu}(a_n\zeta) + \frac{i}{2}{\bm J}_{\nu}(a_n\zeta)
     - \frac{i}{2}{\bm J}_{\nu}(a_n\zeta) \right],
\end{eqnarray}
where~$I_{\nu}(t)$ is the Bessel function of the second kind in the standard notation
and~${\bm J}_{\nu}(t)$ is the Anger function~\cite{GradshteinBook}.

Thus we expressed kernels~$\hK^{\pm}({\bm r}_1,{\bm r}_2)$ by
combinations of~$\delta({\bm r}_1-{\bm r}_2)\hat{1}$,~$\delta({\bm r}_1-{\bm r}_2)\hb$
and~$\hT_0({\bm r}_1,{\bm r}_2)$ operators, see Eq.~(\ref{DK_Upm}).
Operator~$\hT_0({\bm r}_1,{\bm r}_2)$ of Eq.~(\ref{T0_sumn})
can be rewritten in a block form
\begin{equation} \label{FK_T0}
\hT_0({\bm r}_1,{\bm r}_2) = \left(\begin{array}{cc} \hTT_1({\bm r}_1,{\bm r}_2) & \hTT_2({\bm r}_1,{\bm r}_2) \\
                   \hTT_2({\bm r}_1,{\bm r}_2) & -\hTT_1({\bm r}_1,{\bm r}_2) \end{array}\right),
\end{equation}
where
\begin{eqnarray}
\label{FK_T1}
 \hTT_1({\bm r}_1,{\bm r}_2) = \sum_{n=0}^{\infty}
    \left(\begin{array}{cc} D_1 \Lambda_{n-1,n-1} & 0 \\ 0 & D_1 \Lambda_{n,n} \end{array} \right), \\
\label{FK_T2}
 \hTT_2({\bm r}_1,{\bm r}_2) = \sum_{n=0}^{\infty}
    \left(\begin{array}{cc} D_3 \Lambda_{n-1,n-1} & - \sqrt{n} D_2 \Lambda_{n-1,n} \\
     - \sqrt{n}D_2 \Lambda_{n,n-1} & - D_3 \Lambda_{n,n} \end{array} \right).
\end{eqnarray}
Using Eqs.~(\ref{DK_Kpm}) and~(\ref{T_T0}) we also can rewrite
kernels~$\hK^{\pm}({\bm r}_1,{\bm r}_2)$ in a block form
\begin{equation} \label{FK_K22}
\hK^{\pm}({\bm r}_1,{\bm r}_2) = \left(\frac {\hb\pm \hat{1}}{2\sqrt{2}}\right)
   \left(\begin{array}{cc}\hd & \hTT_2 - i\hTT_1 \\
   \hTT_2 + i\hTT_1 & \hd \end{array}\right),
\end{equation}
where we denote~$\hd = \delta({\bm r}_1-{\bm r}_2)\hat{1}$. Then the kernels are
\begin{eqnarray} \label{FK_K44}
\hK^{\pm}({\bm r}_1,{\bm r}_2) = \left(\frac {\hb\pm \hat{1}}{2\sqrt{2}}\right) \times
 \nonumber \\
\hspace*{-9em} \sum_{n=0}^{\infty}
 \left(\begin{array}{cccc}
    \delta({\bm r}_1-{\bm r}_2) & 0 & (D_3-iD_1) \Lambda_{n-1,n-1} & -\sqrt{n} D_2\Lambda_{n-1,n} \\
     0 & \delta({\bm r}_1-{\bm r}_2) & -\sqrt{n} D_2\Lambda_{n,n-1}& -(D_3+iD_1) \Lambda_{n,n} \\
    (D_3+i D_1) \Lambda_{n-1,n-1} & -\sqrt{n} D_2\Lambda_{n-1,n}   & \delta({\bm r}_1-{\bm r}_2) & 0 \\
    -\sqrt{n} D_2\Lambda_{n,n-1} & (-D_3+iD_1) \Lambda_{n,n}  & 0  & \delta({\bm r}_1-{\bm r}_2)
   \end{array}\right). \ \ \ \ \
\end{eqnarray}
Equation~(\ref{FK_K44}) together with Eqs.~(\ref{T0_Lambda_mn}) -~(\ref{T0_Lambda_nn1}),
(\ref{T0_D1}) -~(\ref{T0_D3}) are the final expressions for the matrix elements of kernel
operator. The kernels~$K^{\pm}({\bm r}_1,{\bm r}_2)$ are given by combinations of
two parts: translational-invariant gauge-independent parts
and the gauge-dependent phase factor~$e^{i\chi}$. Since~$\chi$
is not translational invariant, the kernels~$K^{\pm}({\bm r}_1,{\bm r}_2)$
are functions of independent variables~${\bm r}_1$ and~${\bm r}_2$.

The non-diagonal elements of matrix~(\ref{FK_K44}) are non-local in~${\bm r}_1$ and~${\bm r}_2$
coordinates. In consequence, they change shapes of functions transformed
with the use of~$K^{\pm}({\bm r}_1,{\bm r}_2)$, see Eq.~(\ref{DK_Psi_pm_r}).
The transformed functions~$\tilde{\Psi}'^{\pm}({\bm r}_1)$ are more delocalized than the initial
functions~$\Psi({\bm r}_1)$, since the non-diagonal elements of matrix~(\ref{FK_K44})
are characterized by the finite extent on the order of~$\lambda_c$ or of magnetic length~$L$.
We describe this in the next section.

The kernels~$K^{\pm}({\bm r}_1,{\bm r}_2)$ depend on two independent variables, i.e. on
six coordinates. This case is more complicated than the field-free situation, where the
transformation kernels depend on one independent
variable~${\bm R}={\bm r}_1-{\bm r}_2$, see Refs.~\cite{RoseBook,Rusin2011}.
To simplify our analysis, we set~${\bm r}_2={\bm 0}$ and denote~${\bm r}_1={\bm r}$.
Using this simplification the matrix~(\ref{FK_K44}) can be written as
\begin{eqnarray} \label{RK_K44}
\hK^{\pm}({\bm r},{\bm 0}) = \left(\frac {\hb\pm \hat{1}}{2\sqrt{2}}\right) \times \nonumber \\
\hspace*{-6em}
 \left(\begin{array}{cccc}
    \delta({\bm r}) & 0 & \tilde{\Gamma}_3({\bm r}) -i\tilde{\Gamma}_1({\bm r}) & -\tilde{\Gamma}_2({\bm r}) \\
     0 & \delta({\bm r}) & -\Gamma_2({\bm r}) & -\Gamma_3({\bm r})-i\Gamma_1({\bm r}) \\
    \tilde{\Gamma}_3({\bm r})+i \tilde{\Gamma}_1({\bm r}) & -\tilde{\Gamma}_2({\bm r}) & \delta({\bm r}) & 0 \\
    -\Gamma_2({\bm r}) & -\Gamma_3({\bm r})+i\Gamma_1({\bm r}) & 0 & \delta({\bm r}).
   \end{array}\right).
\end{eqnarray}
In the above equation instead of~$D_i$ and~$\Lambda_{m,n}$ integrals we defined
another three functions
\begin{equation} \label{RK_Gamma1}
\Gamma_1({\bm r}) =\frac{e^{-\bRho^2/2+i\chi}}{\pi\lambda_cL^2}
    \sum_{n=0}^{\infty}K_0(a_n|\zeta|){\rm L}_n^0 (\bRho^2), \end{equation}
\begin{equation} \label{RK_Gamma2}
\Gamma_2({\bm r}) =\frac{\eta r_{1,0}e^{-\bRho^2/2+i\chi}}{\pi\lambda_cL^2}
   \sum_{n=1}^{\infty}K_0(a_n|\zeta|) {\rm L}_{n-1}^1 (\bRho^2), \end{equation}
\begin{eqnarray} \label{RK_Gamma3}
\Gamma_3({\bm r}) &=&\frac{ie^{-\bRho^2/2+i\chi}}{\pi\lambda_cL^2}
    \sum_{n=0}^{\infty}\left[\frac{1}{z} + {\cal A}(a_n\zeta)\right]{\rm L}_n^0(\bRho^2) \nonumber \\
    &=& \frac{i\delta(\bRho^2)}{\pi\lambda_cL^2z} +\frac{ie^{-\bRho^2/2+i\chi}}{\pi\lambda_cL^2}
             \sum_{n=0}^{\infty} {\cal A}(a_n\zeta){\rm L}_n^0(\bRho^2). \ \ \ \
\end{eqnarray}
We introduced the notation~$\bRho^2=(x^2+y^2)/(2L^2)$.
In the second line of Eq.~(\ref{RK_Gamma3}) we expressed the infinite sum over Laguerre
polynomials by~$\delta({\bm \rho})$ function, see~\ref{AppDelta}.
Functions~$\tilde{\Gamma}_1({\bm r})$ and~$\tilde{\Gamma}_3({\bm r})$
are defined similarly to~$\Gamma_1({\bm r})$, and~$\Gamma_3({\bm r})$
replacing~$a_n\rightarrow a_{n+1}$. The function~$\tilde{\Gamma}_2({\bm r})$
can be obtained from~$\Gamma_2({\bm r})$ replacing~$r_{1,0}\rightarrow r_{0,1}$, see
Eqs.~(\ref{T0_Lambda_n1n}) -~(\ref{T0_Lambda_nn1}).

The summations over~$n$ in Eqs.~(\ref{RK_Gamma1}) -~(\ref{RK_Gamma3}) extend to infinity
but in practice they are limited to finite numbers of terms by an exponential decrease
of~$K_0(a_n|\zeta|)$ and~${\cal A}(a_n\zeta)$
with~$n$. However, for low magnetic fields there are large numbers of Laguerre polynomials
involved in the summations so we truncate them for~$K_0(a_n|\zeta|)< 10^{-6}$
and~$|{\cal A}(a_n\zeta)|< 10^{-6}$ or, alternatively, for~$n\ge 10^6$.
This approximation allows us to perform calculations of~$\Gamma_i({\bm r})$ for~$B>10^{7}$~T, see below.
For strong magnetic fields there is~$\eta\rightarrow \infty$, so the
arguments~$a_n\zeta=(z/\lambda_c)\sqrt{1+\eta^2n}$ of the Bessel or
Anger functions in Eqs.~(\ref{RK_Gamma1}) -~(\ref{RK_Gamma3}) are large for~$n \neq 0$,
and the terms including~$a_0\zeta$ do not depend on the field. Then the summations over~$n$ reduce to single
terms with~$n=0$. In this limit there is~$\Gamma_2({\bm r})\simeq 0$, while
\begin{eqnarray}
 \label{RK_Gamma1_n0}
\Gamma_1({\bm r}) &\simeq&\frac{e^{-\bRho^2/2+i\chi}}{\pi\lambda_cL^2} K_0(|z|/\lambda_c), \\
 \label{RK_Gamma3_n0}
\Gamma_3({\bm r}) &\simeq&\frac{i\delta(\bRho^2)}{\pi\lambda_cL^2z} +
         \frac{ie^{-\bRho^2/2+i\chi}}{\pi\lambda_cL^2}{\cal A}(z/\lambda_c).
\end{eqnarray}
Thus both~$\Gamma_1({\bm r})$ and~$\Gamma_3({\bm r})$ have the extent~$L$ in the~$x-y$
directions and~$\lambda_c$ in the~$z$ direction, so they are strongly anisotropic.
For~$\tilde{\Gamma}_1({\bm r})$,~$\tilde{\Gamma}_2({\bm r})$ and~$\tilde{\Gamma}_3({\bm r})$
functions there are no field-independent terms in the series (because~$a_0 \rightarrow a_1$),
so all these terms vanish in high magnetic fields and no widening occurs.

\begin{figure}
\includegraphics[width=8.0cm,height=8.0cm]{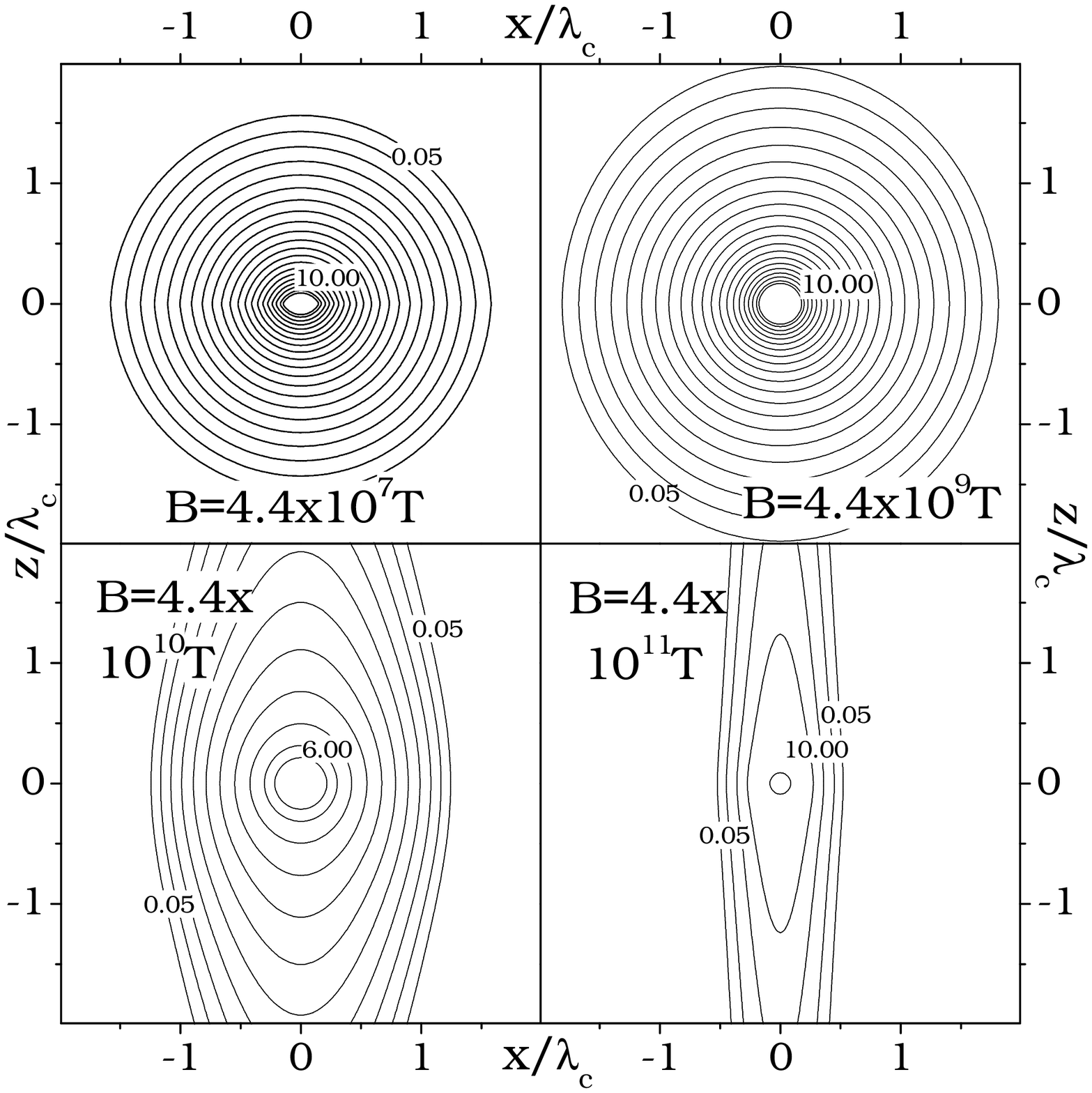}
\caption{
 Calculated contour plot of~$\Gamma_1({\bm r})$, as given in Eq.~(\ref{RK_Gamma1}), for the plane~$y=0$.
 Contour lines start for~$\Gamma_1({\bm r})=0.05\lambda_c^{-3}$ (outer lines)
 and end for~$\Gamma_1({\bm r})=10\lambda_c^{-3}$. Coordinates~$x$ and~$z$ are in~$\lambda_c$ units.
 For low magnetic fields the kernel is almost symmetric in~$x,y,z$ coordinates, while for large~$B$
 the kernel in~$x$ and~$y$ directions is much more confined than in~$z$ direction.} \label{Fig1G1}
\end{figure}

In Figure~\ref{Fig1G1} we show a contour plot of~$\Gamma_1({\bm r})$ function
for the~$y=0$ plane calculated
for various magnetic fields. For each of the four sections the contour lines start
from~$\Gamma_1({\bm r})=0.05\lambda_c^{-3}$ and they increase
logarithmically to~$10\lambda_c^{-3}$.
For all magnetic fields and all~$x$ and~$z$ the
values of~$\Gamma_1({\bm r})$ are positive and they monotonically decrease to zero
with growing~$x$ and~$z$.
For~$z=0$ the expressions for~$\Gamma_1({\bm r})$ diverge [see Eq.~(\ref{RK_Gamma1})],
but for~$x=0$ and nonzero~$z$ they are large but finite.

In Figure~\ref{Fig1G1}a we show~$\Gamma_1({\bm r})$ for a magnetic field~$B=4.4\times 10^7$ T, which
corresponds to~1\% of the Schwinger field of~$B_0=4.4\times 10^9$ T.
For such a field~$\Gamma_1({\bm r})$ is, to a good approximation,
spherically isotropic. The characteristic width of~$\Gamma_1({\bm r})$ is on
the order of~$\lambda_c$ in all directions, as expected for the zero-field limit
of the Dirac equation~\cite{RoseBook}. Similar results are obtained for lower magnetic fields.
At the Schwinger field (Figure~\ref{Fig1G1}b) the
term~$\Gamma_1({\bm r})$ is still almost spherically isotropic but the widening of~$\Gamma_1({\bm r})$
is larger than in the low-field limit. For still higher fields the terms~$\Gamma_1({\bm r})$
become anisotropic: their extent in the~$z$ direction changes slowly but they significantly
shrink in the~$x-y$ directions.
The behavior of~$\Gamma_2({\bm r})$ and~$\Gamma_3({\bm r})$ functions is similar to
that described for~$\Gamma_1({\bm r})$, however~$\Gamma_2({\bm r})$ vanishes in the
limits of~$B\rightarrow \infty$ or~$B\rightarrow 0$.
For magnetic fields lower than the Schwinger field the behavior
of~$\tilde{\Gamma}_j({\bm r})$ terms is similar to the analogous~$\Gamma_j({\bm r})$ terms.
For higher fields all these terms go to zero.

\section{Second moments of transformation kernels}

Now we estimate quantitatively the spatial extent of the elements of kernel
matrices~$\hK^{\pm}({\bm r},{\bm 0})$ given in Eq.~(\ref{RK_K44}) by
calculating second normalized moments~${\cal W}_j$ of~$\Gamma_j({\bm r})$
and~$\tilde{\Gamma}_j({\bm r})$ functions (j=1,2,3).
Here we follow Rose~\cite{RoseBook} who used second moments to characterize non-locality
of the transformation kernel for the Foldy-Wouthyusen transformation in absence of
external fields. The second normalized moments are defined as
\begin{equation} \label{SE_W}
{\cal W}_j = \frac{\int r^2\Gamma_j({\bm r}) d^3{\bm r}}{\int \Gamma_j({\bm r}) d^3{\bm r}}.
\end{equation}
The second normalized moments~$\tilde{\cal W}_j$ of~$\tilde{\Gamma}_j({\bm r})$ functions
are defined similarly.
The above integrals over~$\Gamma_2({\bm r})$,~$\Gamma_3({\bm r})$,~$\tilde{\Gamma}_2({\bm r})$
and~$\tilde{\Gamma}_3({\bm r})$ functions vanish due to symmetry. Below we analyze in more detail
the second moments~${\cal W}_1$ and~$\tilde{\cal W}_1$ of the~$\Gamma_1({\bm r})$
and~$\tilde{\Gamma}_1({\bm r})$ functions.
Introducing auxiliary quantities
\begin{eqnarray}
\label{SE_Mz}
{\cal M}_{\nu}^{z}&=& \frac{1}{\pi\lambda_c} \int_{-\infty}^{\infty} z^{\nu} K_0(|z|a_n/\lambda_c) dz, \\
\label{SE_Mrho}
{\cal M}_{\nu}^{\rho}&=& \frac{1}{L^2} \int_{0}^{\infty} \int_0^{2\pi} e^{-\bRho^2/2 + ixy/(2L^2)}
                          {\rm L}_n^0(\bRho^2) \rho^{\nu+1} d\rho d\phi, \nonumber \\
\end{eqnarray}
where~$\bRho=\rho/(\sqrt{2}L)$, we obtain
\begin{equation} \label{SE_WM}
{\cal W}_1 = \frac{\sum_{n=0}^{\infty} {\cal M}_0^z {\cal M}_2^{\rho} +
             \sum_{n=0}^{\infty} {\cal M}_2^z {\cal M}_0^{\rho}}
             {\sum_{n=0}^{\infty} {\cal M}_0^z {\cal M}_0^{\rho}}
             \equiv {\cal W}_1^{\rho} +{\cal W}_1^z.
\end{equation}
Direct calculations give~${\cal M}_{0}^z=1/a_n$ and~${\cal M}_{2}^z=\lambda_c^2/a_n^3$,
where~$a_n$ are defined in Eq.~(\ref{T0_an}).
On setting in Eq.~(\ref{SE_Mrho}):~$\nu=0$ and~$t=\bRho^2$,
making the substitution~$xy/(2L^2)=\rho^2\sin(2\phi)/(4L^2)$ and
integrating over~$d\phi$, we obtain
\begin{equation} \label{SE_M0rho}
{\cal M}_{0}^{\rho}=\int_0^{\infty} e^{-t/2}J_0(t/2){\rm L}_n^0(t) dt = \sqrt{2}{\rm P}_n(0),
\end{equation}
where~${\rm P}_n(x)$ are the Legendre polynomials. Using the above results, we have
\begin{eqnarray}
\label{SE_M0z0rho}
\sum_{n=0}^{\infty} {\cal M}_0^z {\cal M}_0^{\rho}&=&\sqrt{2}\sum_{n=0}^{\infty}\frac{{\rm P}_n(0)}{a_n}, \\
\label{SE_M2z0rho}
\sum_{n=0}^{\infty} {\cal M}_2^z {\cal M}_0^{\rho}&=&\sqrt{2}\
 \lambda_c^2\sum_{n=0}^{\infty}\frac{{\rm P}_n(0)}{a_n^3}.
\end{eqnarray}
To calculate~${\cal M}_{2}^{\rho}$ we proceed in a similar way and obtain
\begin{equation} \label{SE_M2rho}
{\cal M}_{2}^{\rho}=2L^2\int_0^{\infty} e^{-t/2}J_0(t/2) t{\rm L}_n^0(t) dt.
\end{equation}
In order to use the result~(\ref{SE_M0rho}) we must express the product~${\rm L}_n^0(t)t$
in terms of~${\rm L}_{m}^0(t)$ polynomials. Using the
identities:~$(n+1){\rm L}_{n+1}^0(t)-(2n+1-t){\rm L}_{n}^0(t)+n{\rm L}_{n-1}^0(t)=0$
and~${\rm L}_{0}^0(t)t={\rm L}_{0}^0(t)-{\rm L}_{1}^0(t)$ we have
\begin{eqnarray} \label{SE_M0z2rho}
\hspace*{-7em}
\sum_{n=0}^{\infty} {\cal M}_0^z {\cal M}_2^{\rho}=\sum_{n=1}\frac{2L^2}{a_n}\int_0^{\infty}e^{-t/2}J_0(t/2){\rm L}_n^0(t)t dt
    + \frac{2L^2}{a_0}\int_0^{\infty} e^{-t/2}J_0(t/2){\rm L}_0^0(t)t dt \nonumber \\
\hspace*{-5em}
    = 2L^2\sqrt{2} \left\{\sum_{n=0}^{\infty} {\rm P}_n(0) \left(\frac{1}{a_n}-\frac{1}{a_{n+1}} \right) +
        \sum_{n=0}^{\infty} n{\rm P}_n(0) \left(\frac{2}{a_n}-\frac{1}{a_{n+1}} -\frac{1}{a_{n-1}}\right) \right\},
\end{eqnarray}
where we put~$1/a_{-1}=0$ and used the identities:~${\rm P}_{2n+1}(0)=0$ and~$n{\rm P}_n(0)=0$
for~$n=0,1$. The sums over~$n$ can be calculated numerically either noting that for~$m=2n$
there is~${\rm P}_{m}(0)=(-1/4)^m\left(\begin{array}{c} 2m \\ m \end{array}\right)$ or,
alternatively, by a direct summation with the use of
the generating functions for the Legendre polynomials, see~\ref{AppInt}.
Calculating the second moment of~$\tilde{\Gamma}_1({\bm r})$ one should
substitute~$a_n \rightarrow a_{n+1}$ in
equations~(\ref{SE_M0z0rho}),~(\ref{SE_M2z0rho}) and~(\ref{SE_M0z2rho}).

In the limit of high fields there is~$\eta\gg 1$ and~$1/a_n \simeq \delta_{n,0}$, so
the second moment of~$\Gamma_1({\bm r})$ in the~$x-y$ directions
is~${\cal W}_1^{\rho} \simeq 2L^2\propto B^{-1}$, while
the second moment of~$\Gamma_1({\bm r})$ in the~$z$ direction
is~${\cal W}_1^z \simeq \lambda_c^2$, see Eq.~(\ref{RK_Gamma1_n0}).
For~$\tilde{\cal W}_1$ there are no field-independent terms in the summations in
Eqs.~(\ref{SE_M0z0rho}),~(\ref{SE_M2z0rho}) and~(\ref{SE_M0z2rho}),
and in high fields there is~$\tilde{\cal W}_1^{\rho} \simeq \tilde{\cal W}_1^{z} \simeq 0$.
Thus no widening occurs. For low magnetic fields, when~$\eta \ll 1$, we may
approximate~$a_n^{-p}\simeq 1 - p\eta^2n$.
Since~$\sum_n {\rm P}_n(0)=1/\sqrt{2}$ (see~\ref{AppInt}) we have
from Eq.~(\ref{SE_M0z0rho}) that~$\lim_{B\rightarrow 0}\sum_{n=0}^{\infty} {\cal M}_0^z {\cal M}_0^{\rho}=1$.
This gives, upon using Eqs.~(\ref{SE_WM}),~(\ref{SE_M2z0rho}) and~(\ref{SE_M0z2rho}),
\begin{eqnarray} \label{Lim_Wz0}
\lim_{B\rightarrow 0}{\cal W}_1^{z}&\simeq& \sqrt{2}\lambda_c^2\sum_{n=0}^{\infty}{\rm P}_n(0)
      (1-3\eta^2n/2) \simeq \lambda_c^2, \\
 \label{Lim_Wrho0}
\lim_{B\rightarrow 0}{\cal W}_1^{\rho}
 &\simeq& 2L^2\sqrt{2}\left\{\sum_{n=0}^{\infty}{\rm P}_n(0) \frac{\eta^2}{2}\right\}= 2\lambda_c^2,
\end{eqnarray}
The same results are obtained for~$\tilde{\cal W}_1^{\rho}$ and~$\tilde{\cal W}_1^{z}$ moments.
Thus the second moments of~$\Gamma_1({\bm r})$ and~$\tilde{\Gamma}_1({\bm r})$ functions
in the low-field limit are
\begin{equation} \label{Lim_WTot0}
\lim_{B\rightarrow 0}{\cal W}_1= \lim_{B\rightarrow 0}\tilde{\cal W}_1=3\lambda_c^2.
\end{equation}
This result reproduces our value obtained previously for MOT at~$B=0$, see Ref.~\cite{Rusin2011}.

\begin{figure}
\includegraphics[width=8.0cm,height=8.0cm]{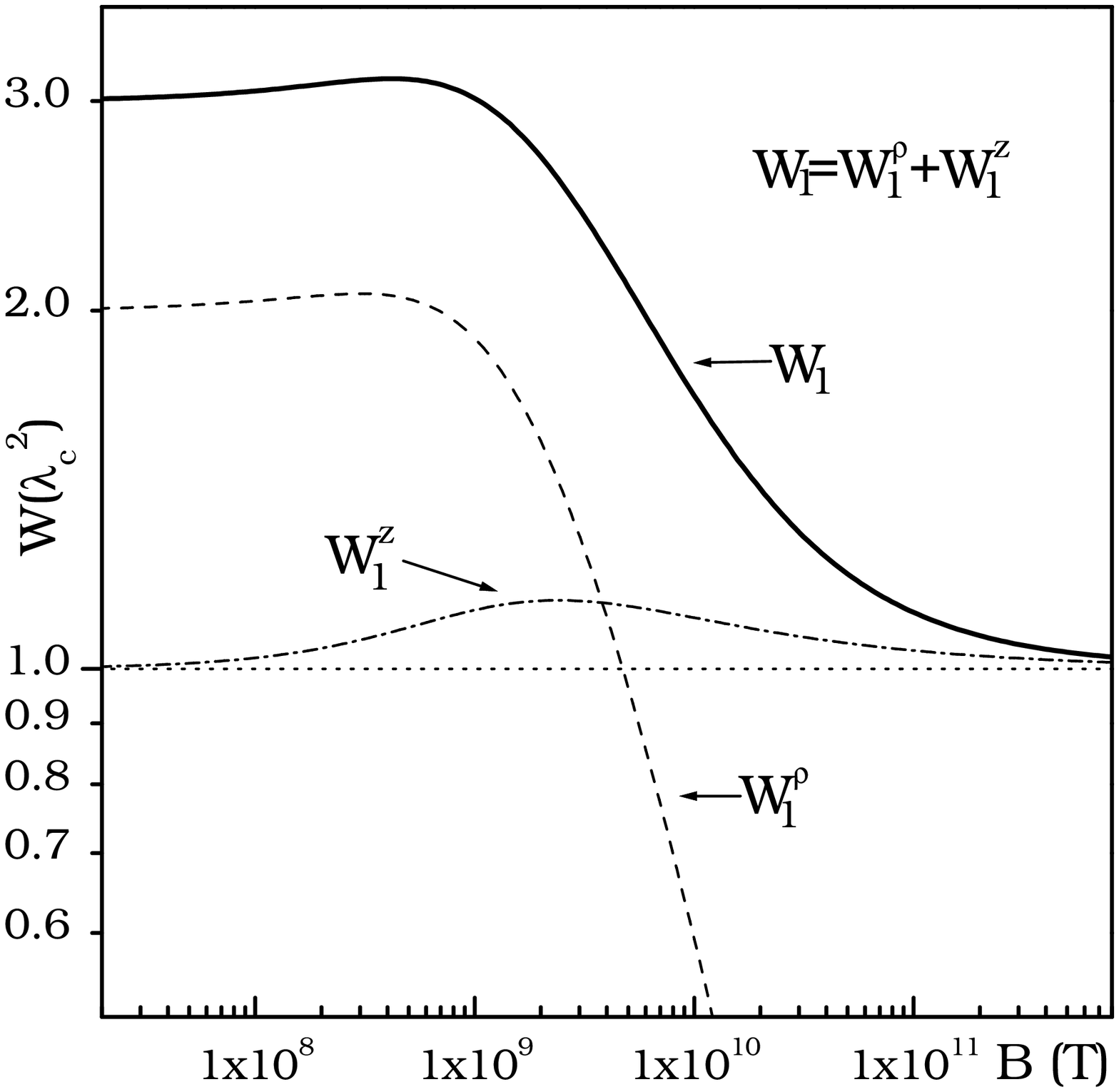}
\caption{
 Calculated normalized second moment of~$\Gamma_1({\bm r})$, as given in Eq.~(\ref{SE_WM}),
 versus magnetic field~$B$.
 Solid line: total second moment~${\cal W}_1$, dashed line:~${\cal W}_1^{\rho}$,
 dot-dashed line:~${\cal W}_1^z$. Second moment~${\cal W}_1$ exhibits a transition
 from 3D case~${\cal W}_1\simeq 3\lambda_c^3$ at low fields to 1D
 case~${\cal W}_1\simeq \lambda_c^2$ at high fields.} \label{Fig2Moment}
\end{figure}

In Figure~\ref{Fig2Moment} we show~${\cal W}_1$ (thick line),~${\cal W}_1^{\rho}$ dashed line)
and~${\cal W}_1^z$ (dash-dotted line),
as defined in Eqs.~(\ref{SE_WM}),~(\ref{SE_M0z0rho}),~(\ref{SE_M2z0rho})
and~(\ref{SE_M0z2rho}), versus magnetic field~$B$.
For low magnetic fields~${\cal W}_1^{\rho}$
tends to~$2\lambda_c^2$ and remains nearly constant up to the Schwinger
field, where a flat maximum occurs. For larger magnetic fields~${\cal W}_1^{\rho}$
starts to decrease and finally decreases as~$B^{-1}$.
The second moment~${\cal W}_1^z$ remains nearly constant,~${\cal W}_1^z \simeq \lambda_c^2$,
within the whole range of magnetic fields except in the vicinity of the Schwinger field.
The total second moment~${\cal W}_1={\cal W}_1^{\rho}+{\cal W}_1^z$ of
the~$\Gamma_1({\bm r})$ function,
exhibits for an increasing magnetic field a transition
from the~3D value~${\cal W}_1 =3\lambda_c^2$ to the~1D
value~${\cal W}_1=\lambda_c^2$. For low magnetic fields~$\tilde{\cal W}_1^{\rho}$
and~$\tilde{\cal W}_1^z$ moments are close to~${\cal W}_1^{\rho}$
and~${\cal W}_1^z$. However, for magnetic fields higher than the Schwinger field both
of them tend to zero exponentially with~$B$. It should be reminded that the kernels given
by Eq.~(\ref{RK_K44}) are composed of various elements. All of them are nonlocal (with the
exception of the delta functions on the diagonal),
but only~$\Gamma_1({\bm r})$ and~$\tilde{\Gamma}_1({\bm r})$ have non-vanishing second moments.
It is the second moment of~$\Gamma_1({\bm r})$ that is shown in Figure~\ref{Fig2Moment}.

For a non-relativistic electron in a magnetic
field there is no widening of the transformed wave function.
We retrieve this limit by setting~$mc^2\rightarrow \infty$,
which gives:~$\lambda_c^2\rightarrow 0$,~$\eta\rightarrow 0$
and~$a_n=1$ for all~$n$. Then the sums in Eq.~(\ref{SE_M0z2rho}) vanish,
the summation in Eq.~(\ref{SE_M0z0rho}) gives unity, and the summation
in Eq.~(\ref{SE_M2z0rho}) gives~$\lambda_c^2 \rightarrow 0$.
Therefore~${\cal W}_1^{\rho}=0$,~$\tilde{\cal W}_1^{\rho}=0$,~$\tilde{\cal W}_1^{z}=0$
and no widening occurs.

\section{Transformed functions}

In this section we consider properties of transformed functions using as an
example a Gaussian function of finite width.
We take the initial function~$\Psi({\bm r})$ in the
form:~$\Psi({\bm r}) = f({\bm r})(0,1,0,0)^T$, in which
\begin{equation} \label{Fun_Gauss}
f({\bm r}) = \frac{e^{-r^2/(2d^2)}}{\pi^{3/4}d^{3/2}},
\end{equation}
and~$d$ characterizes function's width. The transformed function
is:~$\Psi'({\bm r})=\hV\Psi({\bm r}) = f({\bm r})/\sqrt{2}(0,1,0,i)^T$.
Then we have~$|\Psi'^{\pm}\rangle=\hP^{\pm}|\Psi'\rangle$, which gives [see Eq.~(\ref{FK_K44})]
\begin{eqnarray} \label{Fun_Ppm}
\Psi'^{\pm}({\bm r_1}) =
\frac{\sqrt{2}}{2} \left(\begin{array}{c} 0 \\ f({\bm r}_1)\\0 \\if({\bm r}_1) \end{array}\right) \pm
\nonumber \\ \hspace*{-2em}
\frac{\sqrt{2}}{2} \sum_{n=0}^{\infty} \int
\left(\begin{array}{c}
 -i \sqrt{n}D_2(z_1-z_2)\Lambda_{n-1,n}({\bm \rho}_1,{\bm \rho}_2) \\
    \{D_1(z_1-z_2)-iD_3(z_1-z_2)\}\Lambda_{n,n}({\bm \rho}_1,{\bm \rho}_2)\\
     -\sqrt{n} D_2(z_1-z_2)\Lambda_{n-1,n}({\bm \rho}_1,{\bm \rho}_2) \\
  \{iD_1(z_1-z_2)-D_3(z_1-z_2)\}\Lambda_{n,n}({\bm \rho}_1,{\bm \rho}_2)
\end{array}\right) f({\bm r}_2) d^3{\bm r}_2,
\end{eqnarray}
where~$\Lambda_{m,n}$ and~$D_i$ are defined in
Eqs.~(\ref{T0_Lambda_mn}) -~(\ref{T0_Lambda_nn1}) and~(\ref{T0_D1}) -~(\ref{T0_D3}).
As seen from Eq.~(\ref{Fun_Ppm}), positive or negative energy functions~$\Psi'^{\pm}({\bm r_1})$
consist of two parts: the function~$\Psi'({\bm r_1})$ and the nonlocal
parts arising from the integration of~$\Psi'({\bm r_1})$ over appropriate
elements of~$\hT({\bm r_1},{\bm r_2})$. The summation over~$n$ and the threefold integrals
over~${\bm r}_2$ can be performed numerically.

Now we turn to the high-field approximation, in which~$L \ll \lambda_c$.
In this case the summation over~$n$ in Eq.~(\ref{Fun_Ppm}) reduces to the single
term with~$n=0$ and the transformed function is
\begin{equation} \label{Fun_PpmN0}
\Psi'^{\pm}({\bm r}) \simeq
\frac{\sqrt{2C}}{2}
\left(\begin{array}{c} 0 \\ f({\bm r}) \\0 \\ if({\bm r})\end{array}\right) \pm
\frac{\sqrt{2C}}{2} \left(\begin{array}{c} 0 \\ F_z(z)F_{\rho}(\bm \rho)
 \\0 \\i F_z(z)F_{\rho}(\bm \rho) \end{array}\right),
\end{equation}
where~$C$ is a normalization constant and
\begin{equation} \label{Fun_Fz}
F_z(z) =\frac{\sqrt{2d}}{2\pi^{3/4}}\int_{-\infty}^{\infty}
 \frac{(1-i\lambda_ck_z)e^{-k_z^2d^2/2 + ik_zz}}{\sqrt{1+\lambda_c^2k_z^2}} dk_z.
\end{equation}
The presence of Gaussian term ensures the convergence of integration. Further
\begin{eqnarray} \label{Fun_Frho}
F_{\rho}({\bm \rho}) &=& \frac{1}{\pi^{1/2}dL^2} \int e^{-({\bm \rho}-{\bm \rho}_2)^2/4L^2+i\chi}
        e^{-{\bm \rho}_2^2/(2d^2)} d^2 {\bm \rho}_2 \nonumber\\
        &=& \frac{2\sqrt{2\pi}d} {\sqrt{d^4+2d^2L^2+2L^4}}\ e^{-a_x^2x^2-a_y^2y^2+ib^2xy},
\end{eqnarray}
where
\begin{eqnarray}
a_x^2&=&\frac{d^2+L^2}{2(d^4+2d^2L^2+2L^4)}, \\
a_y^2&=&\frac{d^4+d^2L^2+L^4}{2L^2(d^4+2d^2L^2+2L^4)}, \\
b^2  &=&\frac{d^2+L^2}{d^4+2d^2L^2+2L^4}.
\end{eqnarray}
For narrow initial functions:~$d \ll L$, there is~$a_x^2 \simeq a_y^2 \simeq 1/(4L^2)$
and~$b^2\simeq 1/(2L^2)$, so that~$F_{\rho}({\bm \rho})$
has a spatial extent on the order of~$L/2$, similar to the case analyzed before,
see Eqs.~(\ref{RK_Gamma1_n0}) and~(\ref{RK_Gamma3_n0}).
In the opposite limit of wide initial functions:~$d\gg L$,
we have~$a_x^2 \simeq b^2 \simeq 1/(2d^2)$ and~$a_y \simeq 1/(2L^2)$.
The differences between~$a_x$ and~$a_y$
result from the asymmetry of the Landau gauge.
To obtain the same results in another gauge, one would have to modify
the initial functions accordingly, see Refs.~\cite{Kobe1978,Rusin2008}.

To demonstrate a widening of the functions~$\Psi'^{\pm}({\bm r})$ we calculate their
{\it variances}
\begin{equation} \label{Fun_Zpm}
Z^{\pm} = \int \Psi'^{\pm}({\bm r})^{\dagger}{\bm r}^2\Psi'^{\pm}({\bm r}) d^3 {\bm r}
 \left/ \int |\Psi'^{\pm}({\bm r})|^2 d^3 {\bm r} \right.,
\end{equation}
which {\it are different from the second moments} calculated above, see Eq.~(\ref{SE_W}).
The integrals in Eq.~(\ref{Fun_Zpm}) can be calculated analytically
but we do not quote the long resulting expressions.

\begin{figure}
\includegraphics[width=8.0cm,height=8.0cm]{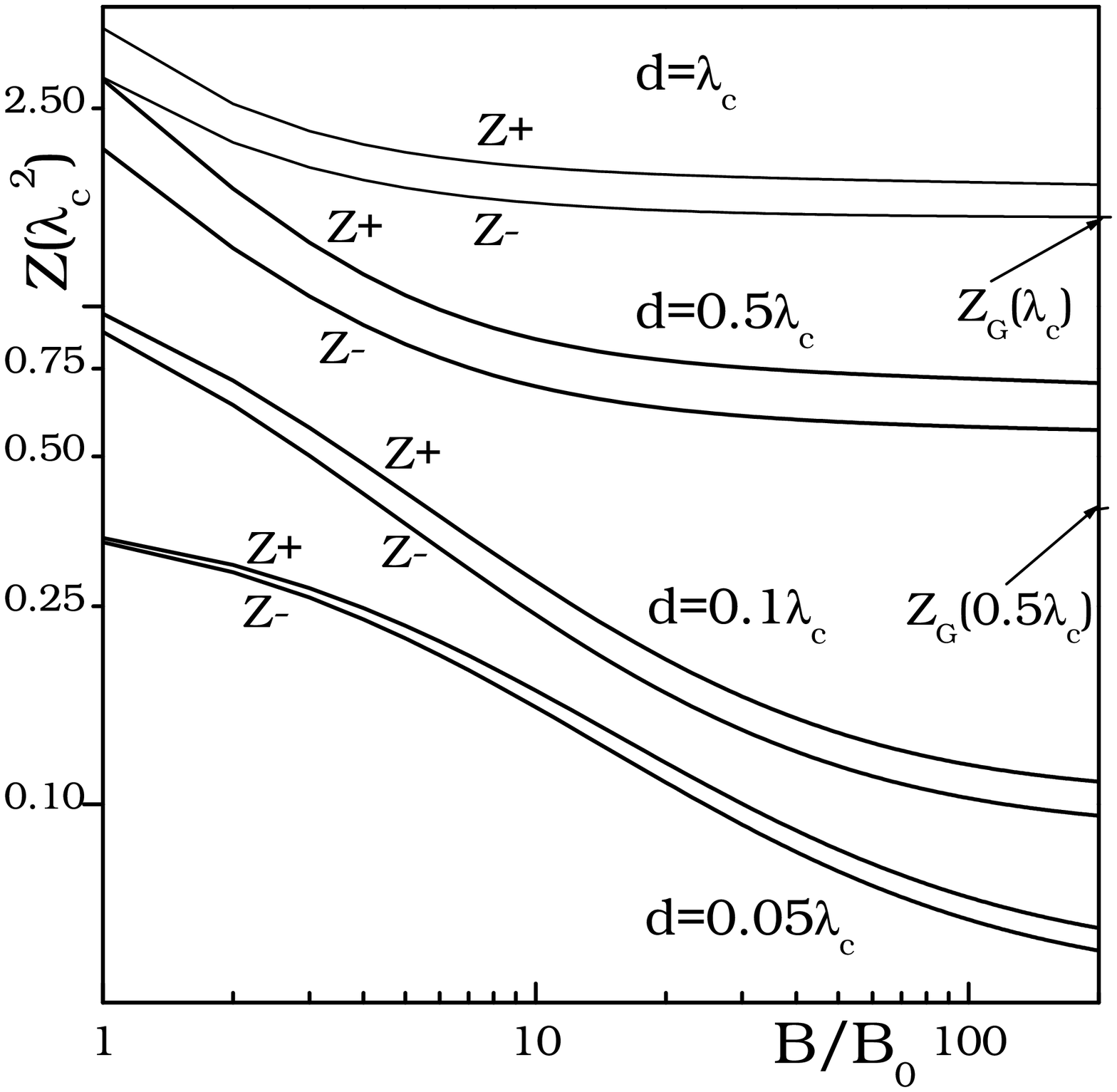}
\caption{
 Calculated variances of transformed Gaussian function~$\tilde{\Psi}'^{\pm}({\bm r})$,
 as given in Eq.~(\ref{Fun_PpmN0}), versus magnetic field~$B/B_0$ in the high-field approximation.
 Solid lines: variances corresponding to the initial functions having different
 widths~$d$. Variances of the initial Gaussian functions for~$d=\lambda_c$ and~$d=0.5\lambda_c$
 are indicated.} \label{Fig3Zpm}
\end{figure}

The plots of~$Z^{\pm}$ for various magnetic fields and initial function widths are
given in Figure~\ref{Fig3Zpm}.
The solid lines represent calculated variances of the transformed function.
The variances~$Z_{G}=3d^2/2$ of the initial functions are indicated on the right ordinate.
For the initial function of the width~$d$ the variances of the transformed functions are larger
than~$Z_{G}$, so the transformed functions are {\it wider} than the initial one.
For the initial functions having~$d=\lambda_c$ the variances~$Z^{\pm}$ reach their limit~$Z_{G}$
for magnetic fields on the order of~$B=10^{11}$ T. For smaller~$d$,
the variances of transformed functions
are much larger than~$Z_{G}$ and they reach the limit~$Z_{G}$ for very large fields.
All results presented in Figure~\ref{Fig3Zpm} are valid for magnetic fields larger
than the Schwinger field, for which~$L \ll \lambda_c$.
The asymmetry between positive and negative variances~$Z^{\pm}$ shown in Figure~\ref{Fig3Zpm}
arises from the asymmetry in the initial wave function having only {\it one} nonzero component.
Selecting other nonzero component or more than one nonzero components, one would obtain different
numerical values of~$Z^{\pm}$, but their
general properties would be similar to those presented in Figure~\ref{Fig3Zpm}.

\section{Two-dimensional case}

\begin{figure}
\includegraphics[width=8.0cm,height=8.0cm]{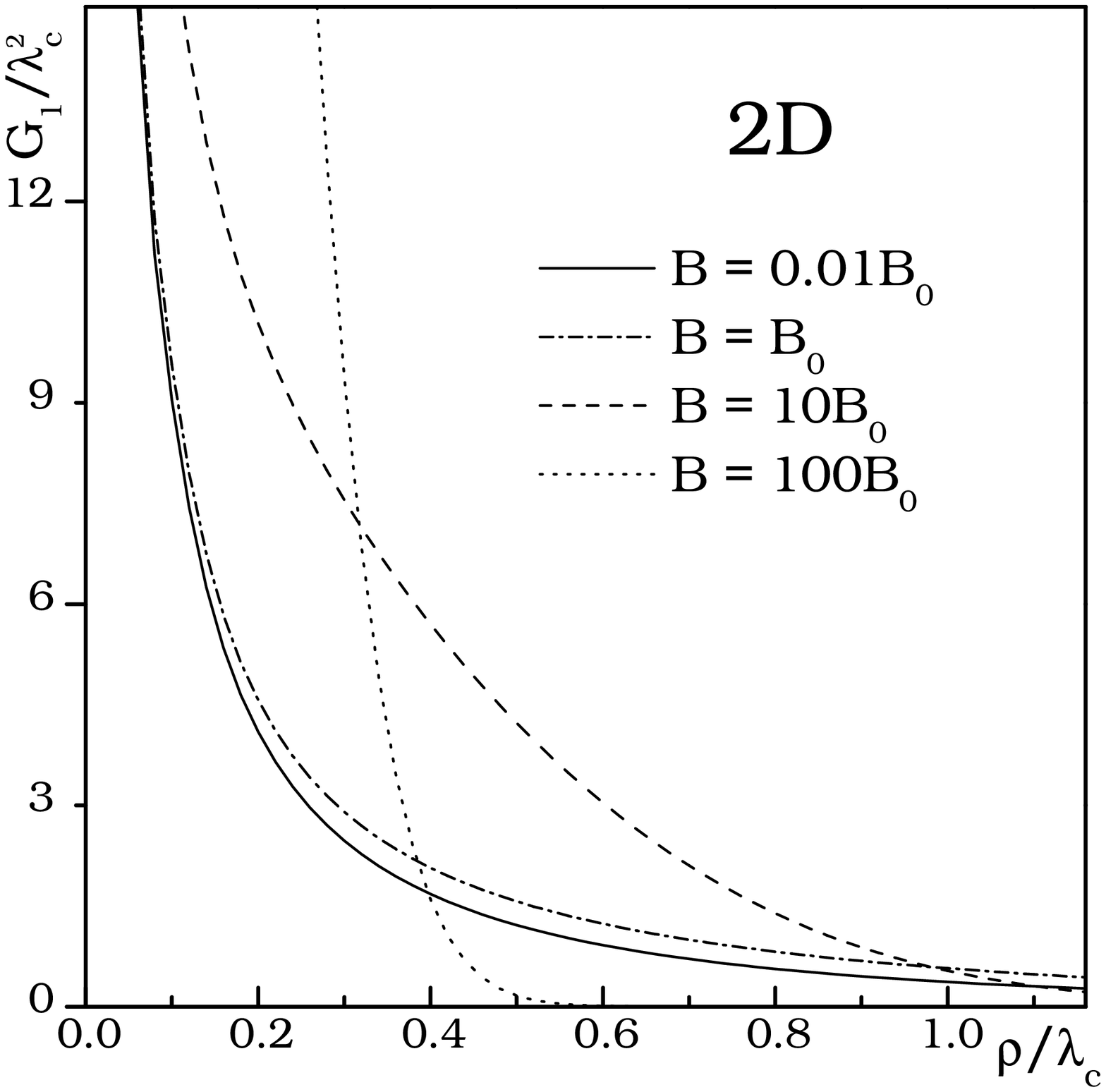}
\caption{Gauge-independent part of kernel's element~$G_1(\rho)$ in two dimensions,
         as given in Eq.~(\ref{RK_G1I}),
         calculated for four values of magnetic field.
         For low fields the kernel non-locality is on the order of~$\lambda_c$ and it does
         not depend on the field. For higher fields the non-locality slightly increases and
         for very large fields it decreases again.} \label{Fig4G12D}
\end{figure}

Recently, Lamata {\it et al.}~\cite{Lamata2011} proposed new topological
effects for the two-dimensional Dirac equation in a magnetic field.
They also indicated a possibility of simulation and
reconstruction of the Wigner function for the lowest Landau levels with the use of
trapped ions. In this connection it is of interest to calculate the non-locality of
the transformation kernel in two dimensions (2D). The matrix elements of 2D kernels
can be obtained from the three dimensional results given above
by setting~$e^{ik_zz} \rightarrow \delta(k_z)$ in all integrals including~$k_z$
and by a proper adjustment of  normalization constants.
The 2D counterpart of Eq.~(\ref{RK_K44}) is
\begin{eqnarray} \label{RK_K44_3D}
\hspace*{-4em}\hK^{\pm}_{2D}({\bm \rho},{\bm 0}) = \left(\frac {\hb\pm \hat{1}}{2\sqrt{2}}\right)
 \left(\begin{array}{cccc}
    \delta({\bm \rho}) & 0 & -i\tilde{G}_1({\bm \rho}) & -\tilde{G}_2({\bm \rho}) \\
     0 & \delta({\bm \rho}) & -G_2({\bm \rho}) & -iG_1({\bm \rho}) \\
    +i \tilde{G}_1({\bm \rho}) & -\tilde{G}_2({\bm \rho}) & \delta({\bm \rho}) & 0 \\
    -G_2({\bm \rho}) & +iG_1({\bm \rho}) & 0 & \delta({\bm \rho}).
   \end{array}\right).
\end{eqnarray}
In the above expression we defined
\begin{equation} \label{RK_G1}
G_1({\bm \rho}) =\frac{e^{-\bRho^2/2+i\chi}}{L^2}
    \sum_{n=0}^{\infty}\frac{{\rm L}_n^0 (\bRho^2)}{\sqrt{1+\eta^2n}}, \end{equation}
\begin{equation} \label{RK_G2}
G_2({\bm \rho}) =\frac{\eta r_{1,0}e^{-\bRho^2/2+i\chi}}{L^2}
   \sum_{n=1}^{\infty}\frac{{\rm L}_{n-1}^1 (\bRho^2)}{\sqrt{1+\eta^2n}}, \end{equation}
using the same notation as in Eq.~(\ref{RK_K44}).
Functions~$\tilde{G}_1({\bm \rho})$ and~$\tilde{G}_2({\bm \rho})$
are defined similarly to~$G_1({\bm \rho})$ and~$G_2({\bm \rho})$
replacing~$\eta^2n\rightarrow \eta^2(n+1)$, see Eqs.~(\ref{T0_Lambda_n1n}) -~(\ref{T0_Lambda_nn1}).
Note that in the 2D case there is no counterpart of the~$\Gamma_3({\bm r})$ term.
Functions~$G_1({\bm \rho})$,~$G_2({\bm \rho})$,~$\tilde{G}_1({\bm \rho})$
and~$\tilde{G}_2({\bm \rho})$ are products of gauge-independent parts
and the gauge-dependent factor $e^{i\chi}$.
The summations over~$n$ in Eqs.~(\ref{RK_G1}) -~(\ref{RK_G2})
are slowly convergent. However, these sums can be transformed to
fast converging integrals that can be calculated numerically.
For~$G_1({\bm \rho})$ one has
\begin{equation} \label{RK_G1I}
G_1({\bm \rho}) =\frac{e^{-\bRho^2/2+i\chi}}{\sqrt{\pi}L^2}
    \int^{-\infty}_{\infty} \frac{1}{1-t}\exp\left(-q^2 +\frac{\bRho^2 t}{t-1}\right) dq,
\end{equation}
where~$t=\exp(-\eta^2 q^2)$. Details of the derivation are given in~\ref{AppInt}.
The remaining sums can be transformed into integrals in a similar way.

In Figure~\ref{Fig4G12D} we plot the gauge-independent part of~$G_1(\rho)$
for four values of magnetic field.
For low fields the spatial extent of the kernel is on the order of~$\lambda_c$
and it does not depend on the field. For the Schwinger field (dash-dotted line)
the non-locality of~$G_1(\rho)$ differs only slightly
from its low-field value (solid line).
For larger fields the non-locality of~$G_1(\rho)$ increases (dashed line) and
for very large fields it decreases again (dotted line). The non-locality
of~$G_1(\rho)$ is on the order of the magnetic length~$L$
which in this regime is much smaller than~$\lambda_c$.
For large fields in three dimensions the non-locality of the kernels tends to zero
in the~$x$ and~$y$ directions, but it remains finite in the~$z$ direction.
For large fields in 2D the non-locality of the kernels goes to zero.
Therefore, for large fields in 2D there is no widening of the transformed
functions. This conclusion may be important for simulations
of the Dirac equation by trapped ions~\cite{Lamata2011}, where arbitrary large values
of a simulated magnetic field may be achieved.

\section{Discussion}

As follows from our considerations and, in particular, from the above figures, the magnetic field
effects for relativistic electrons in a vacuum become appreciable at gigantic field intensities,
comparable to the Schwinger field~$B_0\simeq 4.4\times 10^9$ T. However, one can overcome
this problem in two ways. The first is provided by narrow-gap semiconductors, whose energy band
structure can be described by Dirac-type Hamiltonians with considerably more favorable
parameters, see Ref.~\cite{Zawadzki2005}. For example, in InSb the energy gap,
corresponding to~$2mc^2$ in a vacuum, is~$E_g\simeq 0.23$ eV and the effective mass~$m^*$,
corresponding to rest electron mass~$m$ in a vacuum, is~$m^*\simeq 0.014m$.
In consequence, an effective Schwinger
field in InSb, given by the condition~$\hbar eB_0^*/m^*=E_g/2$, is~$B_0^*\simeq 14$~T.
Such magnetic fields are easily available in laboratories, so the field effects described above
can be readily investigated in narrow-gap semiconductors. One should, however, bear in mind
that in experiments with crystalline solids one deals with additional effects, most notably
with electron scattering leading to broadening of Landau levels, which were not
taken into account in our considerations.

The second way to lower effective magnetic fields is to use simulations of the Dirac equation
employing trapped ions or cold atoms interacting with laser radiation, see e.g.
Refs.~\cite{Leibfried2003,Lamata2007,Johanning2009} and review~\cite{Zawadzki2011}. Such
simulations may also include the presence of an external magnetic fields~\cite{Rusin2010,Lamata2011}.
In this type of experiments an effective simulated field can be tailored to suit
observation requirements. For example, in a simulation of the Dirac equation proposed in
Ref.~\cite{Rusin2010}, the effective simulated magnetic field corresponds to the
ratio~$(\hbar eB/m)/(2mc^2)=16.65$ for realistic trap parameters. However, in simulation-type
experiments there is no need to apply real magnetic fields apart from the fields
necessary to create ion traps.

As to our results, the general rule is that, similarly to the non-locality of energy
separating transformations for~$B=0$, the characteristic quantity is the Compton
wavelength~$\lambda_c=\hbar/mc$, see Ref.~\cite{Rusin2011}. In the presence of a magnetic field,
typical anisotropy appears between the~$z$ direction along the magnetic field and the
transverse~$x-y$ plane. This is seen in the broadening of the transformation kernel
illustrated in Figure~\ref{Fig1G1}. The anisotropy appears also in the second moments
of functional transformation kernels shown in Figure~\ref{Fig2Moment}. This result can
be considered to be a direct extension of that of Rose~\cite{RoseBook}, who characterized
quantitatively the non-locality of the Foldy-Wouthuysen transformation for~$B=0$.
The low field limit~${\cal W}_1=3\lambda_c^2$ agrees with our previous calculation
for MOT at~$B=0$. The corresponding result for FWT is~${\cal W}_1=(3/4)\lambda_c^2$
and one can say that the FW transformation is somewhat more ``compact'' than the MO transformation.
Interestingly, at high fields~${\cal W}_1$ is suppressed, which results
in~${\cal W}_1 \rightarrow \lambda_c^2$. This means that at high~$B$ the quantization
of the electron motion in the~$x-y$ plane gives in consequence an effective
one-dimensional motion along the magnetic field. The parallel second moment~${\cal W}_1^z$
depends somewhat on~$B$ because in the relativistic mechanics there is no exact separation
between parallel and transverse directions of magnetic motion. Finally, Figure~\ref{Fig3Zpm}
indicates two effects. First, the broadening of transformed functions is stronger
for narrow initial widths. Second, with an increasing
magnetic field the broadening of transformed functions diminishes, reaching asymptotically
the non-broadened values at very high fields. One should bear in mind that
the results shown in Figure~\ref{Fig3Zpm} were obtained in the high field approximation,
i.e.~$B/B_0 > 1$.

The results obtained in this paper refer to a non-locality of the energy-separating
transformation for the Dirac Hamiltonian in the presence of a uniform magnetic field.
However, under two assumptions given below it is possible to generalize our approach
to the case of an inhomogeneous magnetic field. First, there should exist an exact
transformation separating the positive and negative energy states.
Second, it is practical to have analytical solutions of the Dirac
equation for electrons in an inhomogeneous magnetic field that allow one to
find an expansion of the kernel into eigenvectors in a way
analogous to that given in Eq.~(\ref{DK_K12}).
More specifically, the existence of transformation separating positive and negative
energy states can be simplified to an assumption that the operator~$\hat{\cal H}^{'2}$
[see Eq.~(\ref{DK_HDp})], describing the square of the Dirac Hamiltonian in an
inhomogeneous magnetic field, is diagonal.
This criterion can be satisfied for example by a vector potential having
the form~${\bm A}=[A_x(x,y),A_y(x,y),0]$ with arbitrary functions~$A_x(x,y)$ and~$A_y(x,y)$.
Then the operator separating the positive and negative energies
is analogous to~$\hU$ given in Eq.~(\ref{DK_U}) but with a suitably changed vector potential.
Physically, the existence of such an operator is limited to situations in which there is no
creation of electron-hole pairs. Therefore, such an operator does not exist for Dirac particle
in an electric field or in some configurations of the magnetic field that mix
positive and negative energy states.
As to the second requirement for the applicability of the present approach,
there are several inhomogeneous magnetic fields known in the literature for which
analytical solutions exist. As an example, in three dimensions the Dirac equation has
solutions for vector potentials~${\bm A}=[(B/a) \tanh (ay),0,0]$
or~${\bm A}=[(B/a)[1 - \exp(ay)],0,0]$, where~$a^{-1}$
is a characteristic length of the potential~\cite{Stanciu1967}.
In two dimensions there exist also solutions for magnetic field~$B=\beta\sec^2(ay)$~\cite{Cooper1995}.
The above vector potentials have the form~${\bm A}=[A_x(x,y),A_y(x,y),0]$,
so that the operator~$\hat{\cal H}^{'2}$ is diagonal and one can calculate the non-locality
of the transformation kernel in a way analogous to that presented above.
We expect the results to be similar to those presented in our work
with properly modified magnetic length.

Our calculations of the non-locality generated
by FWT and MOT at~$B=0$, see Ref.~\cite{Rusin2011}, indicate that different ''separating``
transformations have similar properties and small differences between them
concern mostly the extent of non-localities. In consequence, we believe that the properties
of MOT investigated in this work characterize other separating transformations for the
Dirac electron in the presence of a magnetic field.

\section{Summary}

We investigate non-locality of transformations separating positive and
negative energy states of the Dirac electrons in the presence of a
magnetic field, using as an example the Moss-Okninski transformation.
Functional kernels generated by the transformation are described and their
second moments in the coordinate space are calculated to determine
quantitatively the nonlocal features. The latter are characterized by the
Compton wavelength~$\lambda_c=\hbar/mc$. The behavior of kernels is different
along the magnetic field and in the plane transverse to it. Dependence of
non-locality on the intensity of magnetic field is studied. It is
demonstrated that the transformed functions are broadened. Finally, it is
indicated that a practical way to study Dirac electrons in the
presence of a magnetic field is to simulate their properties using either
narrow-gap semiconductors or trapped ions interacting with the laser
radiation.

\appendix
\section{} \label{AppInt}
Calculating integrals~$\Lambda_{m,n}$ in Eqs.~(\ref{T0_Lambda_mn}) -~(\ref{T0_Lambda_nn1})
we use the identity
\begin{equation} \label{AppInt_HH}
\int_{-\infty}^{\infty}\!\!\!\! {\rm H}_m(q+a) {\rm H}_n(q+b)e^{-q^2}dq =
 2^n\sqrt{\pi}m!b^{n'}{\rm L}_m^{n'}(-2ab),
\end{equation}
where~$n'=n-m$ and~$m\leq n$.

A convenient way to calculate sums in Eq.~(\ref{SE_M0z2rho}) is to use the
generating functions for the Legendre
polynomials:~$\sum_{n=0}^{\infty}t^n {\rm P}_{n}(x)=(1-2tx+t^2)^{-1/2}$
and~$\sum_{n=0}^{\infty}n t^n {\rm P}_{n}(x)=t(x-t)/(1-2tx+t^2)^{3/2}$.
In order to apply the generating functions we first make the
transformation~$1/a_{n}=2/\sqrt{\pi} \int_0^{\infty} \exp[-q^2-q^2\eta^2n]dq$,
next we set~$t=e^{-q^2\eta^2}$, then change the order of summation and integration.
As a result one obtains
\begin{eqnarray} \label{AppInt_Gen1}
\sum_{n=0}^{\infty} \frac{{\rm P}_{n}(0)}{\sqrt{1+\eta^2n}}
 &=& \frac{2}{\sqrt{\pi}}\int_0^{\infty} \frac{e^{-q^2} dq}{\sqrt{1+e^{-2q^2\eta^2}}}, \\
\sum_{n=0}^{\infty} \frac{n{\rm P}_{n}(0)}{\sqrt{1+\eta^2n}}
 &=& \frac{-2}{\sqrt{\pi}}\int_0^{\infty} \frac{e^{-q^2-2q^2\eta^2} dq}{\sqrt{(1+e^{-2q^2\eta^2})^3}}.
\end{eqnarray}
For vanishing~$\eta$ (weak magnetic fields) the above integrals tend to~$1/\sqrt{2}$
and~$-1/\sqrt{8}$, respectively, while for~$\eta \gg 1$ (strong magnetic fields) they tend to~$\pm 1$.

The same method can be used to calculate a sum of the type~$\sum_{n}{\rm P}_{n}(0)/(1+\eta^2n)^{3/2}$,
employing the identity~$1/(1+\eta^2n)^{3/2} = -2(\partial/\partial b)[b+\eta^2n]^{-1/2}$
taken at~$b=1$. Proceeding the same way as for Eq.~(\ref{AppInt_Gen1}) we get
\begin{equation} \label{AppInt_Gen3}
\sum_{n=0}^{\infty} \frac{{\rm P}_{n}(0)}{[1+\eta^2n]^{3/2}} =
\frac{2}{\sqrt{\pi}}\int_0^{\infty} \frac{q^2 e^{-q^2} dq}{\sqrt{1+e^{-2q^2\eta^2}}},
\end{equation}
which applies to Eq.~(\ref{SE_M2z0rho}).

Consider the sum in Eq.~(\ref{RK_G1}) for the 2D case.
Using the identity:~$1/\sqrt{1+\eta^2n}=(1/\sqrt{\pi})\int_{-\infty}^{\infty}e^{-q^2-n\eta^2q^2} dq$
we obtain
\begin{equation} \label{AppInt_S2}
 S =\sum_{n=0}^{\infty}\frac{{\rm L}_n^0 (\bRho^2)}{\sqrt{1+\eta^2n}} =
  \frac{1}{\sqrt{\pi}}\sum_{n=0}^{\infty} \int_{-\infty}^{\infty} {\rm L}_n^0 (\bRho^2)
  e^{-q^2} \left(e^{-\eta^2 q^2}\right)^n.
\end{equation}
Changing the order of summation and integration, and using the identity
\begin{equation}
\sum_{n=0}^{\infty} {\rm L}_n^{\alpha} (x)t^n = \frac{1}{(1-t)^{\alpha+1}}\exp\left(\frac{tx}{t-1}\right)
\end{equation}
with~$t=\exp(-\eta^2 q^2)$ one obtains Eq.~(\ref{RK_G1I}). The presence of a Gaussian term in
Eq.~(\ref{AppInt_S2}) ensures fast convergence of the integral for large~$q$.
For small~$q$ the factor~$\exp[tx/(t-1)]$ tends to zero as~$\exp(-1/q^2)$ and the integrand in
Eq.~(\ref{RK_G1}) remains finite for all values of~$q$.

\section{} \label{AppDelta}
We prove a sum rule for the~$\Lambda_{n,n}$ integrals.
Let~${\bm \rho}_1$ and~${\bm \rho}_2$ be position vectors in the two-dimensional space
and~$\{|{\rm q}\rangle\}$ the complete set of eigenstates of a hermitian operator. Then
\begin{equation} \label{AppDelta_0}
\delta({\bm \rho}_1-{\bm \rho}_2) = \langle {\bm \rho}_1|{\bm \rho}_2\rangle =
\sum_{\rm q}\langle {\bm \rho}_1|{\rm q}\rangle\langle {\rm q}|{\bm \rho}_2\rangle
 = \sum_{\rm q} \psi_{\rm q}({\bm \rho}_1) \psi_{\rm q}({\bm \rho}_2)^{\dagger},
\end{equation}
where~$\psi_{\rm q}({\bm \rho}) = \langle {\bm \rho}|{\rm q}\rangle$.
Taking the eigenstates~$|{\rm q}\rangle=|n,k_x\rangle$ of the Hamiltonian for a
non-relativistic electron in a magnetic field we obtain
\begin{equation} \label{AppDelta_2D}
 \sum_{n=0}^{\infty} \int_{-\infty}^{\infty}\frac{e^{ik_x(x_1-x_2)
      -\frac{1}{2}(\xi_1^2-\xi_2^2)}}{2\pi 2^nn!\sqrt{\pi}L}
  {\rm H}_n(\xi_1) {\rm H}_n(\xi_2)dk_x = \delta({\bm \rho}_1 - {\bm \rho}_2),
\end{equation}
where~${\bm \rho}=(x,y)$.
Performing in Eq.~(\ref{AppDelta_2D}) the integration over~$k_z$ and using the notation
from Eq.~(\ref{T0_Lambda}) we have
\begin{equation} \label{AppDelta_Lambda}
 \sum_{n=0}^{\infty} \Lambda_{n,n}= \delta({\bm \rho}_1 - {\bm \rho}_2).
\end{equation}
Identity~(\ref{AppDelta_Lambda}) was used in Eq.~(\ref{RK_Gamma3}) for the calculation of
infinite sums over the Laguerre polynomials appearing in the~$1/z$ term.

\end{document}